\documentclass[12pt, draftclsnofoot, onecolumn]{IEEEtran}
\usepackage{epsfig,makeidx,color}
\usepackage{amsmath,amssymb,bbm}
\usepackage{cite,graphicx,lipsum}
\usepackage{enumerate}
\def\rT{{\rm T}}

\def\uE{{\mathbb E}}

\DeclareMathOperator*{\argmax}{\arg\!\max}

\newtheorem{mylemma}{\bf Lemma} % [section]
\def\be{ \begin{equation} }
\def\ee{ \end{equation} }
\def\bea{ \begin{eqnarray} }
\def\eea{ \end{eqnarray} }

\def\ba{{\bf a}}

\def\bR{{\bf R}}

\def\b0{{\bf 0}}

\def\cC{{\cal C}}

\def\cB{{\cal B}}

\def\cI{{\cal I}}
\def\cN{{\cal N}}

\def\cT{{\cal T}}

\ifCLASSOPTIONonecolumn
  \newcommand{\figwidth}{0.60\columnwidth}
  
\else
  \newcommand{\figwidth}{0.90\columnwidth}
  
\fi

\begin{document}

\title{Layered Non-Orthogonal Random Access with SIC and 
Transmit Diversity for Reliable Transmissions}

\author{Jinho Choi
\thanks{The author is with School of Electrical
Engineering and Computer Science,
Gwangju Institute of Science and Technology (GIST),
Korea (Email: jchoi0114@gist.ac.kr).
This work was supported by
the ``Climate Technology Development and Application"
research project (K07732) through a grant provided by GIST in 2017.}
}

\maketitle
\begin{abstract}
In this paper, we study a layered random access
scheme based on non-orthogonal multiple access (NOMA)
to improve the throughput of multichannel ALOHA.
At a receiver, successive interference cancellation (SIC)
is carried out across layers to remove the signals that are already decoded.
A closed-form expression for the total throughput is
derived under certain assumptions. It is shown that
the transmission rates of layers can be optimized
to maximize the total throughput and 
the proposed scheme can improve the throughput
with multiple layers.
Furthermore, it
is shown that the optimal rates can be recursively found using
multiple individual one-dimensional optimizations.
We also modify the proposed layered random 
access scheme with contention resolution repetition 
diversity for reliable transmissions with a delay constraint.
It is shown to be possible to have a low outage probability
if the number of copies can be optimized,
which is desirable for high reliability low latency communications.
\end{abstract}


\ifCLASSOPTIONonecolumn
\baselineskip 26pt
\fi

\section{Introduction}

In order to support massive connectivity in 
machine-type communications (MTC), random access has been 
considered in cellular standards \cite{3GPP_MTC} \cite{3GPP_NBIoT}.
In particular, ALOHA 
\cite{Abramson70} \cite{BertsekasBook}
is mainly studied for random access in MTC
with multiple channels, which is multichannel ALOHA \cite{Shen03}.
In \cite{Arouk14} \cite{Chang15},
the performance of multichannel ALOHA has also
been analyzed and optimized for MTC.
In \cite{Choi16}, it is shown that
the number of channels
can be adaptively decided to maximize the throughput 
if the number of channels is flexible. 

Successive interference cancellation
(SIC) can be employed to improve the throughput
when a tree algorithm is used for random access as in
\cite{Yu07}. 
In \cite{Casini07}, within a frame,
a packet is repeatedly transmitted to exploit 
contention resolution repetition diversity (CRRD)
together with SIC.
If there is one copy of a packet that can be transmitted
without collision, it can be successfully decoded and its copies
can be removed. This process
can be repeated at a receiver, which can result 
in the throughput improvement.
In \cite{Liva11} \cite{Paolini15IT}\cite{Paolini15},
further improvements are made using graph-based analysis
for irregular repetition of coded packets.
The resulting approach is referred to as
irregular repetition slotted ALOHA (IRSA).

While the approaches in \cite{Yu07}
assume a simple channel model
without taking into account fading,
it might be necessary to consider fading channels
in random access over wireless channels.
In this case, SIC can be considered in the context of
non-orthogonal multiple access (NOMA) where the power
difference is to be exploited \cite{Choi14}
\cite{Ding14} \cite{Ding_CM}.
In \cite{Choi_JSAC}, a NOMA based 
random access method is proposed where each user 
can randomly choose a channel as well as a power level.
In this case, although two users choose the same channel,
a receiver (or base station (BS) for uplink
transmissions) can recover both signals 
using SIC if they choose different power levels.
Thus, the throughput can be improved.
In \cite{Liang17}, NOMA is also used for random access
based on the power control scheme proposed in \cite{Zhang16}.

In this paper, we study a NOMA based random access
as in \cite{Choi_JSAC} to improve
the throughput of multichannel ALOHA by
exploiting the power difference.
The resulting random access scheme can be seen as
a layered random access scheme where each layer
is characterized by the power level.
However, unlike \cite{Choi_JSAC},
we assume that users do not know their channel state information (CSI).
We derive a closed-form expression for the throughput
in terms of transmission rates under Rayleigh fading channels.
This closed-form expression allows us to find the optimal rates
that maximize the total throughput. 

Although the proposed random access scheme in this paper relies on
SIC as IRSA, there are a few key differences.
The proposed scheme does not use iterative decoding, which is used in
\cite{Liva11} \cite{Paolini15IT}\cite{Paolini15}. Thus, the decoding delay 
at a receiver is fixed.
In terms of interference cancellation,
the main difference is that the 
proposed scheme uses inter-layer interference cancellation,
which is SIC across layers. On the other hand, IRSA
uses intra-layer interference cancellation
(as there is only one layer),
where
interference cancellation is repeatedly carried out until no more
signals are decodable.
Furthermore, along with SIC,
the proposed scheme exploits the capture effect,
which is considered in \cite{Goodman87} \cite{Zorzi97}
to exploit the near/far effect,
while IRSA does not consider the capture effect.
In this paper, the capture effect is induced by the power difference
for NOMA.
%While only the strongest signal is to be recovered
%the capture effect is considered in random access 
%in packet capture, SIC could recover more signals.

We also consider a modification of the proposed scheme
with CRRD where multiple copies of a packet 
are transmitted through randomly selected different channels.
Unlike the approaches in \cite{Liva11} \cite{Paolini15IT}\cite{Paolini15},
the main aim of this modification (with CRRD) is to guarantee
a packet delivery 
subject to a delay constraint
with a high probability, not to improve the throughput.
However, as there are multiple layers, the overall throughput can be
reasonably high once SIC is successfully used.
The resulting approach may be useful 
for high
reliable low latency communications \cite{Johansson15}\cite{Durisi16}.

The rest of the paper is organized as follows.
In Section~\ref{S:SM},
we present the system model for random access over fading channels
and introduce the proposed layered random access scheme.
We study the throughput of 
the proposed scheme in Section~\ref{S:TA},
where the proposed scheme is also briefly compared with IRSA.
In Section~\ref{S:RC}, the proposed scheme is modified with
random CRRD for reliable transmissions with a delay constraint.
Simulation results are shown in Section~\ref{S:SM} and the paper
is finally concluded with some remarks in Section~\ref{S:Con}.

\subsubsection*{Notation}
Matrices and vectors are denoted by upper- and lower-case
boldface letters, respectively.
The superscript $\rT$
denotes the transpose.
The 2-norm is denoted by $||\ba||$.
$\uE[\cdot]$ and ${\rm Var}(\cdot)$
denote the statistical expectation and variance, respectively.
$\cC\cN(\ba, \bR)$
represents the distribution of
a circularly symmetric complex
Gaussian (CSCG) random vector with mean vector $\ba$ and
covariance matrix $\bR$.

\section{System Model}	\label{S:SM}

In this section, we present the system model
for layered random access and explain a receiver algorithm
that is based on SIC.

\subsection{Layered Random Access}

Suppose that a system consists of one BS\footnote{Throughout
this paper, the BS and receiver are interchangeable as we consider
uplink random access.}
and a number of users for random access with
multiple channels in a time slot, 
which might be equivalent to a multiple access control (MAC) frame
in \cite{Liva11} \cite{Paolini15IT}\cite{Paolini15}.
To improve the throughput, we consider
multiple layers for each channel by
exploiting the notion of NOMA.

We assume that there are $L$ layers and each layer,
which is characterized by a different power level, consists
of $N$ orthogonal channels\footnote{We can use orthogonal 
frequency division multiple access (OFDMA) 
to form multiple orthogonal 
channels in a time slot.}
in a time slot.
An active user that has a packet to transmit is to randomly choose
a channel and one of $L$ layers in the selected channel.
Let $\cI_{l,q}$ denote the 
index set of the users that choose channel $q$ and layer $l$.
In addition, denote by $h_{k,q}$ and $d_k$
the channel coefficient and data symbol from user $k$ to the BS,
respectively, provided that user $k$ chooses channel $q$.
Throughout the paper, we assume block-fading channels \cite{TseBook05} where 
the channel coefficients, $h_{k,q}$, remain unchanged within a time slot.
Then, the received signal at the BS through channel $q$ is given by
\be
y_q = \sum_{l=1}^L s_{l,q} + n_q,
\ee
where $n_q \sim \cC \cN(0, N_0)$ is the background noise and
$s_{l,q} =
\sum_{k \in \cI_{l,q}} \sqrt{P_l} h_{k,q} d_k$.
Here, $P_l$ is the transmit power 
of layer $l$, which is a design parameter.
Here, we assume that $\uE[|d_k|^2] = 1$ and $\uE[d_k] = 0$.
If a user chooses layer $l$, the signal power has to be set to $P_l$.

Note that if $L = 1$, the resulting system becomes
the conventional single-channel slotted ALOHA system. 
Throughput the paper, we assume coded signals from users.
If a user chooses layer $l$, the transmission (or code)
rate is set to $R_l$. Together with $P_l$,
$R_l$ is also a design parameter.

\subsection{Signal Decoding using SIC}

Let 
\be
x_{l,q} = s_{l,q} + n_{l,q},
\ee
where $n_{l,q}$ is the interference (plus-noise) term in layer $l$,
which is given by
$$
n_{l,q} = \sum_{i=l+1}^L s_{i,q} + n_q.
$$
Clearly, $x_{1,q} = y_q$
and $n_{L,q} = n_q$.
If there are multiple signals in $x_{l,q}$,
we may assume that the receiver cannot decode any signal 
due to packet collision. However, 
if there is only one signal and the interference
is sufficiently weak in layer $l$
(provided that the signals in layers $1, \ldots
l-1$ are removed), the receiver can decode the signal.
Let $\beta_l$ denote the conditional probability of decoding error 
at layer $l$ when there is only one signal from user $k$ at 
channel $q$ under the lower-interference-free condition.
Then, at channel $q$, assuming that capacity achieving codes
are used, we have
\be
\beta_l = 
\Pr \left( \log_2 
\left( 1 + \frac{P_l |h_{k,q}|^2 }{\sigma_{l,q}^2} \right) < R_l \right),
	\label{EQ:beta_l}
\ee
where $\sigma_{l,q}^2 
= {\rm Var}(n_{l,q}\,|\,\{h_{k,q}\})
= \uE[|n_{l,q}|^2 \,|\,\{h_{k,q}\}]$
is the conditional variance of $n_{l,q}$.

At each channel, 
from layer 1 to layer $L$, the receiver performs
decoding with SIC. If there is no signal or one signal
that can be decoded and subtracted, the receiver can move to the next layer
in each channel.
However, if there is packet collision or
a signal that cannot be decoded at a certain layer,
the receiver stops SIC.

\begin{figure}[thb]
\begin{center}
\includegraphics[width=\figwidth]{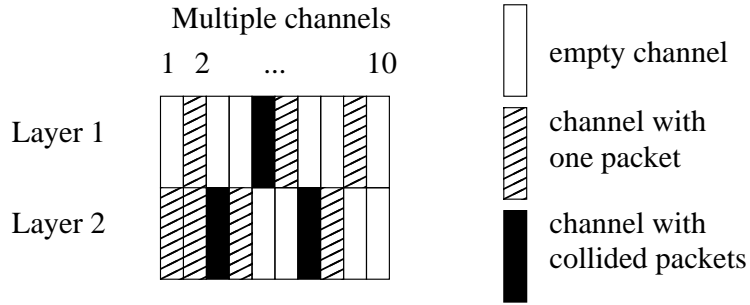}
\end{center}
\caption{Multiple channels of layered random access 
access with $L = 2$ layers and $N = 10$ channels (per layer).}
        \label{Fig:layered}
\end{figure}

In Fig.~\ref{Fig:layered}, we illustrate a set of the
channels of layered random
access with $L = 2$ layers and $N = 10$ channels (per layer).
At channel 1, there is no signal in the first layer,
but in the second layer. Thus, the signal in the second layer
is decodable if the signal-to-noise ratio (SNR) is sufficiently high.
At channel 2, there are two signals. However, they are in different
layers. Thus, the BS is able to decode the signal in the first layer
if the signal-to-interference-plus-noise ratio (SINR) is sufficiently high.
Once the signal in the first layer is decoded, it can be removed using
SIC and the signal in the second layer can be decoded if its
SNR is sufficiently high.
At channel 3, there are also two signals, but both are in the 
second layer, which results in packet collision. Thus, the two
signals may not be decodable.

In \cite{Choi_JSAC}, by exploiting the power difference,
NOMA is applied to multichannel ALOHA, which becomes the layered
random access scheme in this section.
While it is assumed that the users know their CSI
so that they can decide the transmit powers to allow SIC and guarantee
successful decoding if there is no collision in 
\cite{Choi_JSAC}, we do not assume
CSI at transmitters (i.e., users) in this paper.
Consequently, the success of SIC depends on both packet collision
and (instantaneous) CSI, and the throughput analysis becomes more involved
than that in \cite{Choi_JSAC}.
In the next section, we study the performance analysis under certain
assumptions.

\section{Throughput Analysis}	\label{S:TA}

We consider the following assumptions in this section for the 
throughput analysis.
\begin{itemize}
\item[{\bf A1}] The number of arrivals at layer $l$ or 
the number of the users 
that choose layer $l$, denoted by $M_l$,
is an independent Poisson random variable with mean $\lambda_l$,
which is the (average) arrival rate\footnote{In MTC,
it is expected to have sparse user activity. As a result,
the normalized arrival rate rate (i.e.,
the arrival rate per channel), $\lambda_l/N$, might be low due
to sparse activity.}
at layer $l$,
i.e.,
\begin{align*}
\Pr(M_l = m) 
& = P_l (m) \cr
& = \frac{\lambda_l^m e^{-\lambda_l}}{m!}, \ m = 0,1,\ldots
\end{align*}
Note that 
the average number of users becomes $\sum_{l=1}^L \lambda_l$. Thus,
if the average number of users is fixed, the
average number of users in each layer, $\lambda_l$, can decrease with $L$.
%{\color{red}
%Note that $\lambda_l$ is also a design parameter that is to be considered
%to improve the throughput.}
\item[{\bf A2}] An active user is to uniformly and randomly choose 
a layer and a channel within the selected layer.
\item[{\bf A3}] If multiple users choose the same channel 
in the same layer, the receiver is not able to recover any signals
and declares packet collision.
\end{itemize}

\subsection{Derivation of Throughput}

Consider layer $l$ and assume that all the signals in the
lower layers, i.e., from layer 1 to layer $l-1$, are successfully
decoded.
Then, under the assumption
of {\bf A3}, the conditional probability of collision or decoding error 
for given $m$ transmitted signals in layer $l$ is given by
\be
\alpha_l (m) = p_{\rm c} (m)
+ \left(1 - p_{\rm c} (m) \right) \beta_l,
	\label{EQ:al}
\ee
where $p_{\rm c} (m)$ is the 
(conditional) probability of packet collision at layer $l$
for given $m$ transmitted signals in layer $l$.
Under {\bf A2}, we have
\be
p_{\rm c} (m) = 1 - \left(1 - \frac{1}{N} \right)^{m-1},
	\label{EQ:A2}
\ee
which is independent of $l$.

\begin{mylemma}	\label{L:1}
Let $\eta_l$ denote the average number of 
successfully decoded packets at layer $l$ provided
that any signals in layers 1 to $l-1$ are removed by SIC.
For convenience, this condition is referred to as
\emph{the lower-interference-free condition} for layer $l$.
Then, under the assumption of {\bf A1}, we have
\be
\eta_l = \varphi_l \lambda_l e^{- \frac{\lambda_l}{N}},
	\label{EQ:eta_l}
\ee
where $\varphi_l = 1 - \beta_l
= 
\Pr \left( \log_2 
\left( 1 + \frac{P_l |h_{k,q}|^2 }{\sigma_{l,q}^2} \right) \ge R_l \right)$.
\end{mylemma}
\begin{IEEEproof}
See Appendix~\ref{A:1}.
\end{IEEEproof}

If $L = 1$, the throughput is given by
\begin{align}
\cT = R_1 \eta_1  
= R_1 \varphi_1 \lambda_1 e^{- \frac{\lambda_1}{N}}.
	\label{EQ:eta1}
\end{align}
The dimension of the throughput 
is the same as that of $R_1$, which is the bits per channel use.
Note that $\lambda_1 e^{- \frac{\lambda_1}{N}}$
is often regarded as the throughput of multichannel ALOHA
\cite{Shen03} (which is the average number of 
packets without packet collision)
under the ideal collision channel model
\cite{BertsekasBook}.

Let $\rho_l$ denote
the probability that there is no transmitted signal 
or \emph{a} transmitted signal is decoded
through a given channel in layer $l$ under the lower-interference-free
condition.
Then, for a given channel,
provided that there are $M_l = m$ signals,
the conditional probability that there is no transmitted signal
is $\left(1 -\frac{1}{N} \right)^m$
and the 
conditional probability that there is a transmitted signal which
is decodable is
$(1 - \beta_l)
\binom{m}{1} \frac{1}{N}
\left( 1 - \frac{1}{N} \right)^{m-1}$.
Then, $\rho_l$ can be found by taking the expectation
over $M_l$, which is given by
\begin{eqnarray}
\rho_l 
& = & \uE \left[
\left(1 - \frac{1}{N} \right)^{M_l}
+ \varphi_l 
\binom{M_l}{1} \frac{1}{N}
\left( 1 - \frac{1}{N} \right)^{M_l-1}
\right] \cr
& = & \sum_{m=0}^\infty 
\left( \left(1 - \frac{1}{N} \right)^m + 
\frac{\varphi_l m}{N}
\left( 1 - \frac{1}{N} \right)^{m-1}
\right) \cr
& & \quad \times P_l (m) \cr
& = & \left( 1 +  \frac{ \varphi_l \lambda_l}{N}
\right) e^{-\frac{\lambda_l}{N}}.
\end{eqnarray}
Suppose that $L = 2$.
Since the numbers of users in layers 1 and 2 are independent, 
the throughput of layer 2 becomes $R_2 \rho_1 \eta_2$.
Thus, the total throughput 
(with $L = 2$) becomes 
$R_1 \eta_1 + R_2 \rho_1 \eta_2$.
Unfortunately, this total throughput is an approximation
since the throughput of each layer is correlated.
To see this clearly, with $L = 2$, we can consider the case
that users $1$ and $2$ send signals through the first channel
in different layers. In this case, 
the probability that the BS can decode both signals becomes
$$
\varphi_{1,2} = \Pr
\left(
\frac{P_1 |h_{1,1}|^2} {P_2 |h_{2,1}|^2 + N_0}\ge \nu(R_1),
\frac{P_2 |h_{2,1}|^2}{N_0} \ge \nu(R_2) \right),
$$
where $\nu (R) = 2^R - 1$.
Clearly, $\varphi_{1,2} \neq \varphi_1 \varphi_2$.
However, if we assume that the events of successful decoding
in different layers are independent (provided that there is only one
signal in each layer at the same channel), 
the throughput can be approximated by
$R_1 \eta_1 + R_2 \rho_1 \eta_2$.
In general, for $L \ge 1$, the total throughput can be approximated 
by $\cT$ that is given by
\begin{align}
\cT & = R_1 \eta_1 + R_2 \rho_1 \eta_2 + 
\ldots + R_L \left(\prod_{m=1}^{L-1} \rho_m \right) \eta_L \cr
& = \sum_{l=1}^L R_l \left( \prod_{m=1}^{l-1} \rho_m \right) \eta_l.
	\label{EQ:T_L}
\end{align}
Throughout the paper, we will consider the approximate
throughput in \eqref{EQ:T_L} and, for convenience,
$\cT$ is simply
referred to as the throughput (although it is an approximation).
Note that in \eqref{EQ:T_L}, $\prod_{m=1}^{l-1} \rho_m$ 
is the probability that 
the lower-interference-free condition holds at layer $l$
(under the independence assumption).

\subsection{Throughput Maximization}

In this subsection, we consider the throughput maximization.
As shown in \eqref{EQ:T_L}, in order to maximize
the throughput, the power and rate allocation as well
as the arrival rate control can be considered.
However, due to tractability, we only focus on 
the rate optimization to maximize
the throughput, while the arrival rates
and powers are fixed.

For the throughput maximization, we consider the following assumption
for fading channels.
\begin{itemize}
\item[{\bf A4}] The channel coefficients, $h_{k,q}$, are iid and $|h_{k,q}|$
is Rayleigh distributed with $\uE[|h_{k,q}|^2] = \sigma_h^2$.
\end{itemize}

\begin{mylemma}	\label{L:2}
Under the assumptions of {\bf A1} -- {\bf A4},
we have
\begin{align}
\varphi_l & = 
\exp \left(- \frac{\nu(R_l) N_0}{ P_l \sigma_h^2}
- \sum_{i=l+1}^L \frac{\lambda_i }{N} \frac{ 
\nu(R_l) P_i}{P_l + \nu(R_l) P_i} \right) \cr
& \ge  \exp \left( - \frac{\nu(R_l)}{\gamma_l} \right),
	\label{EQ:1b}
\end{align}
where 
$\gamma_l$ is the target SINR at layer $l$, which is
given by
$\gamma_l = \frac{P_l \sigma_h^2}{\bar \sigma_{l}^2}$.
Here,
\be
\bar \sigma_l^2 = \sum_{i = l+1}^L \frac{\sigma_h^2 P_i \lambda_i}{N}
+ N_0.
	\label{EQ:bsl}
\ee
\end{mylemma}
\begin{IEEEproof}
See Appendix~\ref{A:2}.
\end{IEEEproof}

In \eqref{EQ:1b}, $\varphi_l$
can be seen as the capture probability 
\cite{Goodman87} \cite{Zorzi97}, which is the probability
that the receiver
can decode the signal in layer $l$ in the presence of
the interferences in the upper layers,
i.e., layers $l+1,\ldots, L$.
In general, we can see that $\varphi_l$
is a function of $P_l, \ldots, P_L$,
$\lambda_{l+1}, \ldots, \lambda_L$, and $R_l$.
Assuming that $\lambda_l$ and $P_l$ are fixed or given,
$\varphi_l$ becomes a function of $R_l$,
the throughput, $\cT$,
is a function of the rates, $\{R_1, \ldots, R_L\}$
and the throughput maximization is given by
\be
\{R_l^*, \ l = 1,\ldots, L\} =
\argmax_{R_l \ge 0, \ l = 1,\ldots, L} \cT(R_1, \ldots, R_L),
	\label{EQ:RT}
\ee
which is an $L$-dimensional optimization
problem. 
Fortunately, it is not necessary
to perform a high dimensional optimization to find the optimal
rates in \eqref{EQ:RT}.
For example, consider the case of $L = 2$, where the throughput
is given by
\begin{align}
\cT (R_1, R_2) 
= R_1 \eta_1 (R_1) + R_2 \eta_2 (R_2) \rho_1 (R_1),
	\label{EQ:T2}
\end{align}
which suggests that the throughput can be maximized by finding the optimal
value of $R_2$ first and then that of $R_1$ for given optimal rate
$R_2^*$.
Based on this, we can see that the optimal rates
to maximize the total throughput can
be found by $L$ individual one-dimensional optimizations
as follows.

\begin{mylemma}	\label{L:opt_rate}
Each optimal transmission rate can be
individually and recursively found in descending order as
\be
R_l^* = \argmax_{R_l} T_l (R_l), \ l= L, L-1, \ldots, 1,
	\label{EQ:Rest}
\ee
where 
\begin{align}
T_l (R_l) 
= R_l \eta_l (R_l) + \rho_l (R_l) T_{l+1} (R_{l+1}^*)
	\label{EQ:bT}
\end{align}
and $T_L (R_L) = R_L \eta_L (R_L)$.
\end{mylemma}
\begin{IEEEproof}
It is a straightforward generalization from \eqref{EQ:T2},
we omit the proof.
\end{IEEEproof}

Note that 
the transmit powers can be decided to keep the average SINR of each
layer constant as follows:
$$
\frac{P_l \sigma_h^2}{\bar \sigma_l^2} = \gamma, 
$$
where $\gamma$ represents the target SINR, i.e.,
$P_l$ is in descending order decided as
\be
P_l = \frac{\gamma \bar \sigma_l^2}{\sigma_h^2}, \ l = L, \ldots, 1,
	\label{EQ:Pl}
\ee 
since $\bar \sigma_l^2$ is a function of 
$P_{l+1},\ldots, P_L$ as shown in \eqref{EQ:bsl}.

Note that each user is to randomly choose a layer 
as all the users have the same average channel power gains 
under the assumption of {\bf A4}. 
This assumption differs from that
in \cite{Zhang16} \cite{Liang17},
where each user may have a different average channel power gain. 
As in \cite{Zhang16} \cite{Liang17},
if users have different average channel power gains and 
know them, each user 
can choose the layer according to the average channel power gain
so that the overall transmit power is minimized.
For example,  if $L = 2$, 
the users in a cell
can be divided into two groups depending on their distances
from the BS. A group of users 
whose distances from the BS are less than or equal
to $d_{\rm in}$, which is a threshold
distance and less than cell radius, can be called
near users, while the other users whose distances are greater than
$d_{\rm in}$ can be called far users.
Since a near user 
can have a higher channel gain than a far user with the same transmit power,
it might be reasonable to allocate 
near users to layer 1 and far users to layer 2 to reduce the transmit power.
With $L > 2$, this approach can be straightforwardly generalized.
%and a similar approach is considered in 
%\cite{Choi_JSAC}.

\subsection{Comparison with IRSA}

In this subsection, we briefly study
the comparison 
between the proposed layered random access scheme
and IRSA in \cite{Liva11}
in terms of the average number of successfully decoded packets
in a slot (or MAC frame).

In general, the comparison 
between the proposed layered random access scheme
and IRSA in \cite{Liva11}
is not straightforward since different assumptions 
are used in each scheme (e.g.,
no fading is considered in \cite{Liva11}).
However, under some additional 
assumptions and approximations, we can consider comparisons as follows.

For the sake of simplicity,
we consider the lower-bound in \eqref{EQ:1b}
and a fixed target SINR, $\gamma$.
If we assume that $R = R_l$ for all $l$, then 
$\varphi_l \ge \tilde \varphi = 
e^{- \frac{\nu(R) }{\gamma}}$,
which is assumed to be fixed.
In this case, 
$\eta_l (R_l)$ is lower-bounded by 
$$
\tilde \eta_l (R_l) = \tilde \eta_l =
\tilde \varphi \lambda_l e^{-\frac{\lambda_l}{N}} 
$$
and $\rho_l$ is also lower-bounded by
$\tilde \rho_l = \left( 1+ \frac{\tilde \varphi \lambda_l}{N}
\right) e^{-\frac{\lambda_l}{N}}$.
Then,
the average number of 
successfully decoded packets
is lower-bounded 
by
\begin{align}
S_{\rm LA} & \ge 
\tilde S_{\rm LA} =
\sum_{l = 1}^L 
\left( \prod_{m=1}^{l-1} \tilde \rho_m \right) \tilde \eta_l \cr
& = N \left(\sum_{l = 1}^L 
\left( \prod_{m=1}^{l-1} \tilde \rho_m \right) \tilde \varphi
\tau_l  e^{-\tau_l} \right),
	\label{EQ:SLA}
\end{align}
where $\tau_l = \frac{\lambda_l}{N}$ is the normalized arrival rate.
In \eqref{EQ:SLA}, we can see that 
$\tilde S_{\rm LA}$ can be maximized 
by finding the optimal arrival rates or normalized arrival rates,
$\{\tau_l\}$. 
Since the optimization can be similar
to that in Lemma~\ref{L:opt_rate}, we do not further discuss it.
However, it is noteworthy that
since $\tilde \rho_l$ is also a function of $\tau_l$
as $\tilde \rho_l = (1+ \tilde \varphi \tau_l) e^{-\tau_l}$,
the average number of successfully decoded packets of the proposed 
scheme can grow linearly with $N$ as shown in \eqref{EQ:SLA}, 
which is similar to 
multichannel ALOHA \cite{Shen03} and IRSA \cite{Liva11}.

\begin{figure}[thb]
\begin{center}
\includegraphics[width=\figwidth]{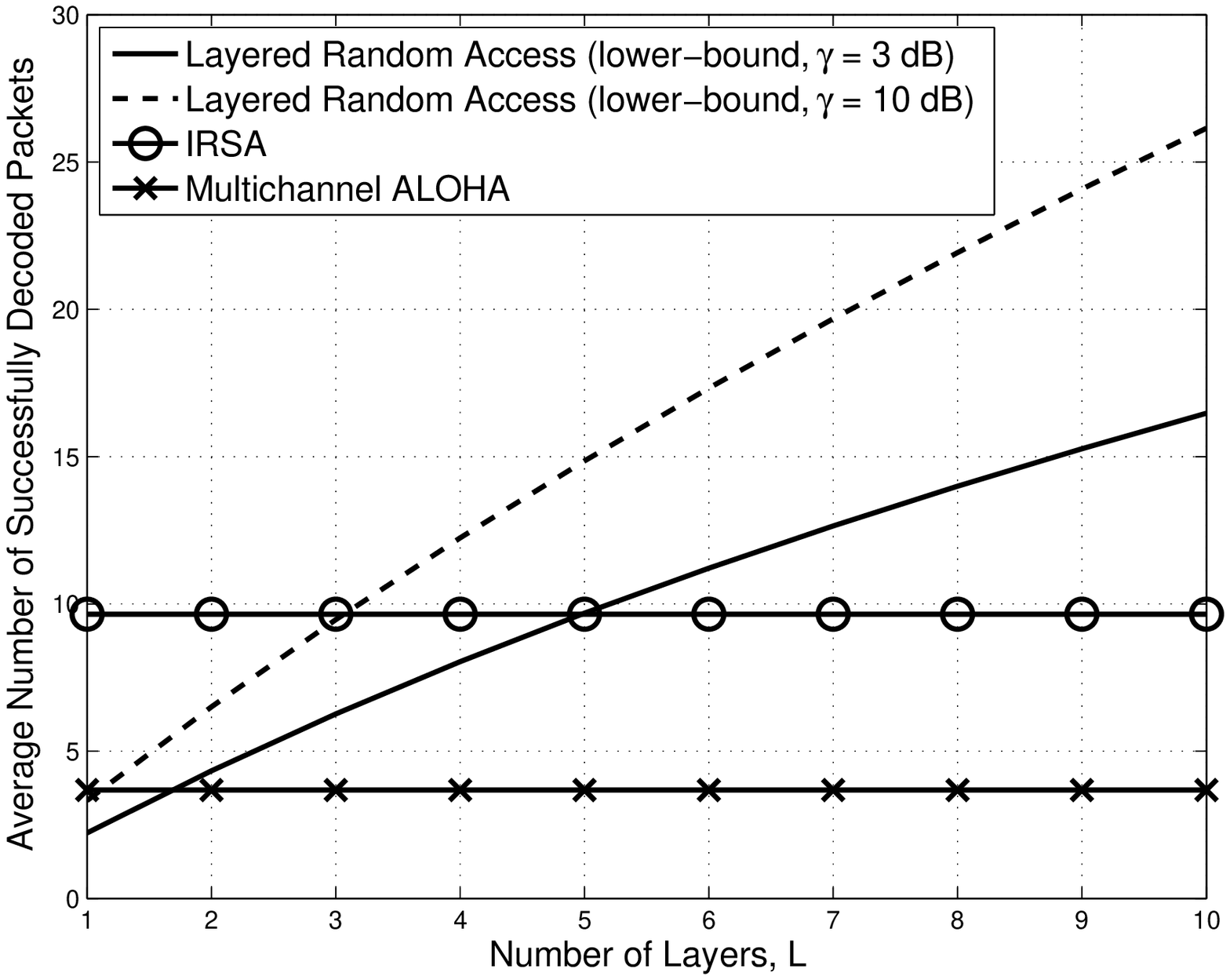} \\
(a) \\
\includegraphics[width=\figwidth]{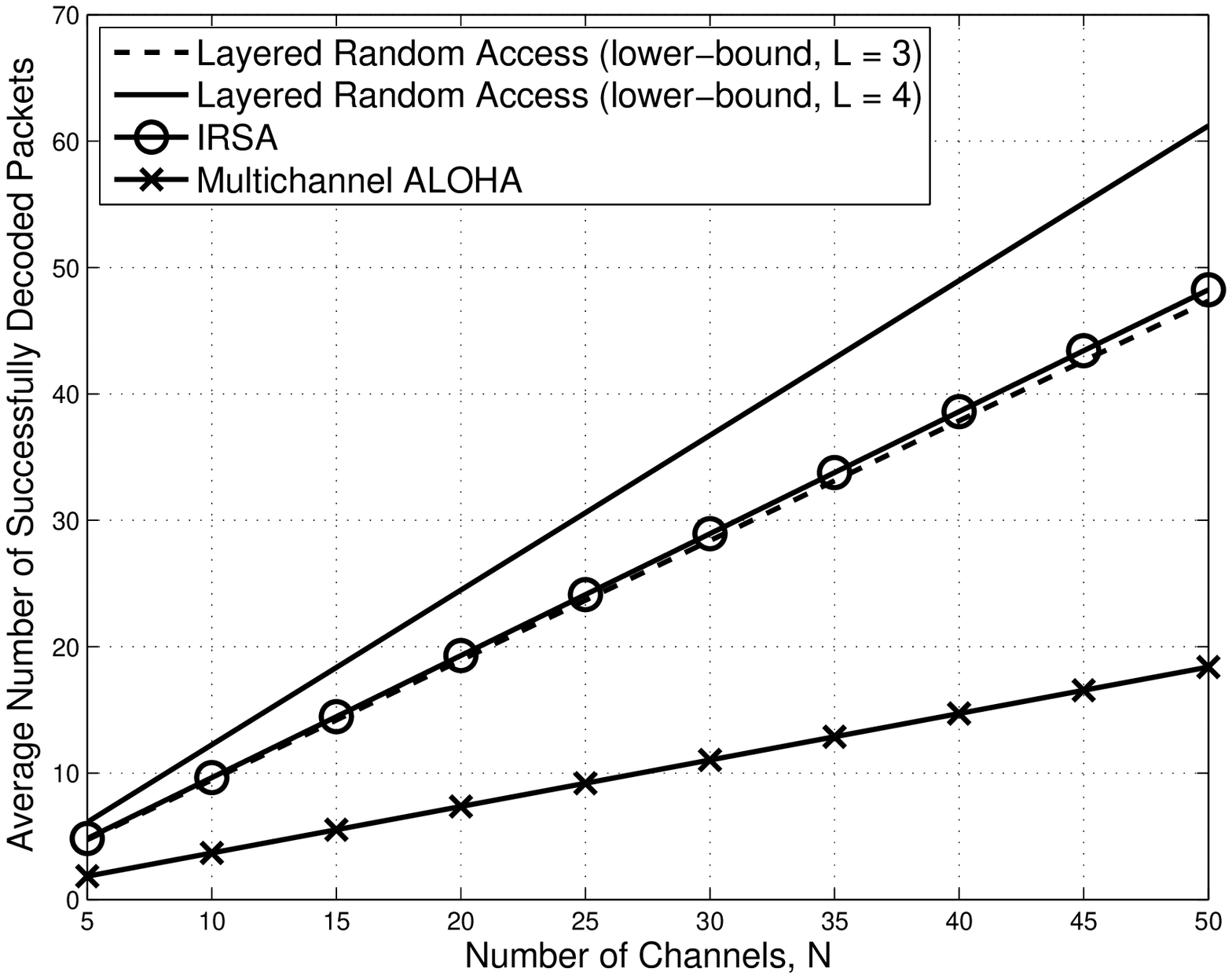} \\
(b) \\
\end{center}
\caption{Comparisons between the proposed layered random access
scheme, IRSA, and multichannel ALOHA in terms of 
the average number of successfully decoded packets: 
(a) 
the average number of successfully decoded packets versus $L$
with $N = 10$;
(b)
the average number of successfully decoded packets versus $N$
with $\gamma = 10$ dB and $L \in \{3, 4\}$.}
        \label{Fig:Tplt}
\end{figure}

Fig.~\ref{Fig:Tplt},
shows the average numbers of successfully decoded packets
for the layered random access scheme\footnote{The arrival
rates are optimized to maximize $\tilde S_{\rm LA}$
in \eqref{EQ:SLA}.}, IRSA\footnote{It is assumed that
the degree distribution is optimized as in \cite{Liva11}
and the asymptotic performance is considered.},
and slotted ALOHA. As mentioned earlier,
for comparisons, we use the lower-bound for the layered random 
access scheme in \eqref{EQ:SLA} with $R = 1$.
In Fig.~\ref{Fig:Tplt} (a), we can see that
the average number of successfully decoded packets
of the layered random access scheme increases with the number of
layers, $L$, while 
the average numbers of successfully decoded packets
of IRSA and multichannel ALOHA are fixed,
which are given by $0.965 N$ (this is obtained from \cite{Liva11}
with a maximum  repetition of 16 as an asymptotic performance)
and $e^{-1}N$ (this is known in \cite{Shen03}
as the maximum stable throughput), respectively,
as they are independent of $L$.

As expected, in 
Fig.~\ref{Fig:Tplt} (b),
it is shown that the 
average numbers of successfully decoded packets
of the layered random access scheme,
IRSA, and multichannel ALOHA grow linearly with $N$.
The proposed layered random access scheme
can provide a higher throughput than the others if $L$
is sufficiently large (e.g., $L \ge 4$).

Note that the comparisons in Fig.~\ref{Fig:Tplt} may not be 
complete as no fading is considered for IRSA\footnote{The performance of 
IRSA under fading is not well studied yet.}
and multichannel ALOHA, while the results in Fig.~\ref{Fig:Tplt} might be 
favorable to IRSA and multichannel ALOHA
(as no fading is considered for them).
In addition, it is noteworthy that the proposed layered random access
scheme can exploit the notion of IRSA within each layer with iterative
decoding. In this case, a receiver can employ
not only inter-layer, but also intra-layer SIC.
This generalization might be a further research topic to be studied
in the future.

\section{Random CRRD for Reliable Transmissions
with a Delay Constraint}	\label{S:RC}

In the previous sections, we have considered the layered random access
scheme that can improve the throughput. This scheme can
be modified to guarantee successful
packet delivery within a slot with a sufficiently high
probability. To this end,
we can consider random CRRD 
where a packet from a user is to be transmitted through randomly selected
multiple channels.

Throughout this section, we consider the following assumption that
replaces {\bf A2}.
\begin{itemize}
\item[{\bf A5}]
A user can transmit $B$ copies of a packet
through randomly selected 
$B$ channels out of $N$ channels, where $B \le N$,
in a randomly (and uniformly) selected layer.
To avoid self-packet collision, 
a user is to choose $B$ different channels.
Each copy of a packet has the pointers of the other $B-1$ copies
as in \cite{Liva11} so that any successfully decoded copy 
can help remove the other copies by interference cancellation.
\end{itemize}

For convenience, $B$ is referred to as the repetition gain.
Clearly, if $B = N$, the resulting random access
is identical to single-channel ALOHA with $N$-fold transmit diversity,
which can be seen as a generalization of \cite{Choudhury83}.
Under the assumption of
{\bf A5}, provided that there are $m$ active users in a layer,
the conditional probability of collision becomes
\be
p_{\rm c} (m) = 1 - \left(1 - \frac{1}{N} \right)^{B (m-1)},
	\label{EQ:A5}
\ee
which can be seen a generalization of \eqref{EQ:A2},
as \eqref{EQ:A5} becomes \eqref{EQ:A2} with $B = 1$. 
Let $\cB_{k,l}$ denote the index set of the selected $B$ channels
by user $k$ provided that user $k$ chooses layer $l$.
The conditional probability of collision or decoding error
of the signal transmitted by a user 
through channel $q \in \cB_{k,l}$,
provided that there are $m$ transmitted signals in layer $l$, becomes
\be
\alpha_l (m) = p_{\rm c} (m) +  (1-p_{\rm c} (m) ) \beta_l,
\ee
where $\beta_l$ is 
the conditional probability of decoding error
of layer $l$ with only one signal from user $k$ at 
channel $q$
under the lower-interference-free condition for layer $l$
and the assumption of {\bf A5}. Note that 
$p_{\rm c} (m)$, 
$\alpha_l (m)$, and $\beta_l$ are slightly different
from those in the previous sections due to multiple transmissions
(or the assumption of {\bf A5}). However, as long as
there is no risk of confusion, we will use the same notations in this
section.

Since there are $B$ copies,
under the assumption of {\bf A3},
the conditional
probability of collision or decoding error
becomes
$\alpha_l^B (m)$
if the signals in channels are independent. Thus,
the (conditional) probability of collision or decoding error
of a user's signal in layer $l$ under the lower-interference-free
condition is given by
\be
\Psi_{l} 
= \sum_{m=1}^\infty \alpha_l^B(m) \bar P_l (m),
	\label{EQ:Psi}
\ee
where
\begin{align}
\bar P_l (m_l) 
= \Pr(m_l \,|\, m_l \ge 1) 
= \frac{P_l (m)}{1 - e^{-\lambda_l}}, \ m_l = 1,\ldots.
\end{align}
Here, $m_l$ represents the number of active users at layer $l$.
Clearly, if $B$ is large, $\alpha_l^B(m)$ is small, which
can result in a low probability of collision or decoding error.
Thus, for a large $B$,
a successful packet transmission subject to 
the delay constraint to transmit within one slot or a MAC frame
can be guaranteed
with a high probability for reliable low latency transmissions
at the cost of throughput.
However, due to multiple layers, it might be possible to 
transmit more packets within a slot.

In fact, \eqref{EQ:Psi}
is valid only if 
the $\sigma_{l,q}^2$'s, $q \in \cB_{k,l}$, are  independent.
%the events that $\frac{P_l |h_{k,q}|^2}{\sigma_{l,q}^2}$,
%$l \in \cB_{l,q}$, are independent.
However,
since each active user sends
$B$ copies of signals to $B$ different channels, 
$\sigma_{l,q}^2$ and $\sigma_{l,q^\prime}^2$ 
can be correlated if any active user in layer $i \in \{l+1, \ldots, L\}$
sends his/her signal to channels $q$ and $q^\prime$ too.
As a result, \eqref{EQ:Psi} is an approximation unless
$B = 1$.
In general, if $B \ll N$,
\eqref{EQ:Psi} might be a reasonably good approximation.

\begin{mylemma}	\label{L:4}
Under the assumptions of {\bf A1}, {\bf A3}, {\bf A4}, and {\bf A5},
the approximation of $\Psi_l$ can be expressed by a sum of finite terms
as follows:
\begin{align}
\Psi_l
& \approx \sum_{m=1}^\infty \alpha_l^B(m) \bar P_l (m) \cr
& = 
\sum_{b = 0}^B  \binom{B}{b}
(1 - \beta_l)^b \beta_l^{B-b} g_l(b),
	\label{EQ:psi_L4}
\end{align}
where 
\begin{align}
g_l(b)  & =
\frac{e^{- \lambda_l} }{1- e^{- \lambda_l} }
\sum_{j=0}^b \binom{b}{j}
(- \omega)^{-j} (e^{\lambda_l \omega^j} - 1) \cr
\beta_l & = 
1 - e^{- \frac{\nu(R_l) N_0}{ P_l \sigma_h^2}
- \sum_{i=l+1}^L \frac{\lambda_i B }{N} \frac{ 
\nu(R_l) P_i}{P_l + \nu(R_l) P_i} }.
	\label{EQ:g_b}
\end{align}
Here, $\omega = \left( 1 - \frac{1}{N} \right)^B$.
\end{mylemma}
\begin{IEEEproof}
See Appendix~\ref{A:4}.
\end{IEEEproof}

From the $\Psi_l$'s, we can define the outage probability of each layer
as the probability that any of $B$ copied packets from
a user cannot be
successfully decoded at layer $l$. The outage probability
of layer 1 is equal to $\Psi_1$. 
For layer $l$, $l > 1$, by taking into account the error
propagation, we have
\be
P_{{\rm out}, l} = 1 - \prod_{i=1}^{l} (1 - \Psi_i).
	\label{EQ:Pel}
\ee
Due to the closed-form expression in 
\eqref{EQ:psi_L4}, the outage probability can be found in a closed-form,
which can help optimize key parameters such as $B$
to minimize the outage probability.
Note that if $\Psi_l \ll \Psi_1$ for $l = 2, \ldots, L$,
we can see that the outage probability is generally
decided by $\Psi_1$ or $P_{{\rm out}, 1}$.
For a low $\Psi_1$, we expect to have reliable transmissions
within a slot.
Furthermore, since $P_{{\rm out}, l}$ increases with $l$,
the users who can accept tolerable reliability
can choose upper layers, i.e., layers $l \in \{2, \ldots, L\}$.

Note that unlike IRSA, no intra-layer SIC  
is used in the modified layered scheme with CRRD in this section.
Furthermore, as no iterative
decoding is used, the processing delay at a receiver is fixed,
which is desirable for high reliability low latency communications.

\section{Simulation Results}
\label{S:Sim}

In this section, we present simulation results
to see the performance of the proposed layered random access
scheme Rayleigh fading in terms of
the throughput (without CRRD) and outage probability (with CRRD).

\subsection{Throughput}

In this subsection, we present simulation results
under the assumptions of {\bf A1} -- {\bf A4}.
For simulations, we assume that the transmit powers are decided
as in \eqref{EQ:Pl} and $\sigma_h^2 = N_0 = 1$ for normalization.
In addition, for convenience, we assume
that $\lambda = \lambda_l$, $l = 1,\ldots,L$
(i.e., an equal arrival rate is assumed for all the layers).
The throughput in this section is in the number of bits per channel 
use as mentioned earlier.

\begin{figure}[thb]
\begin{center}
\includegraphics[width=\figwidth]{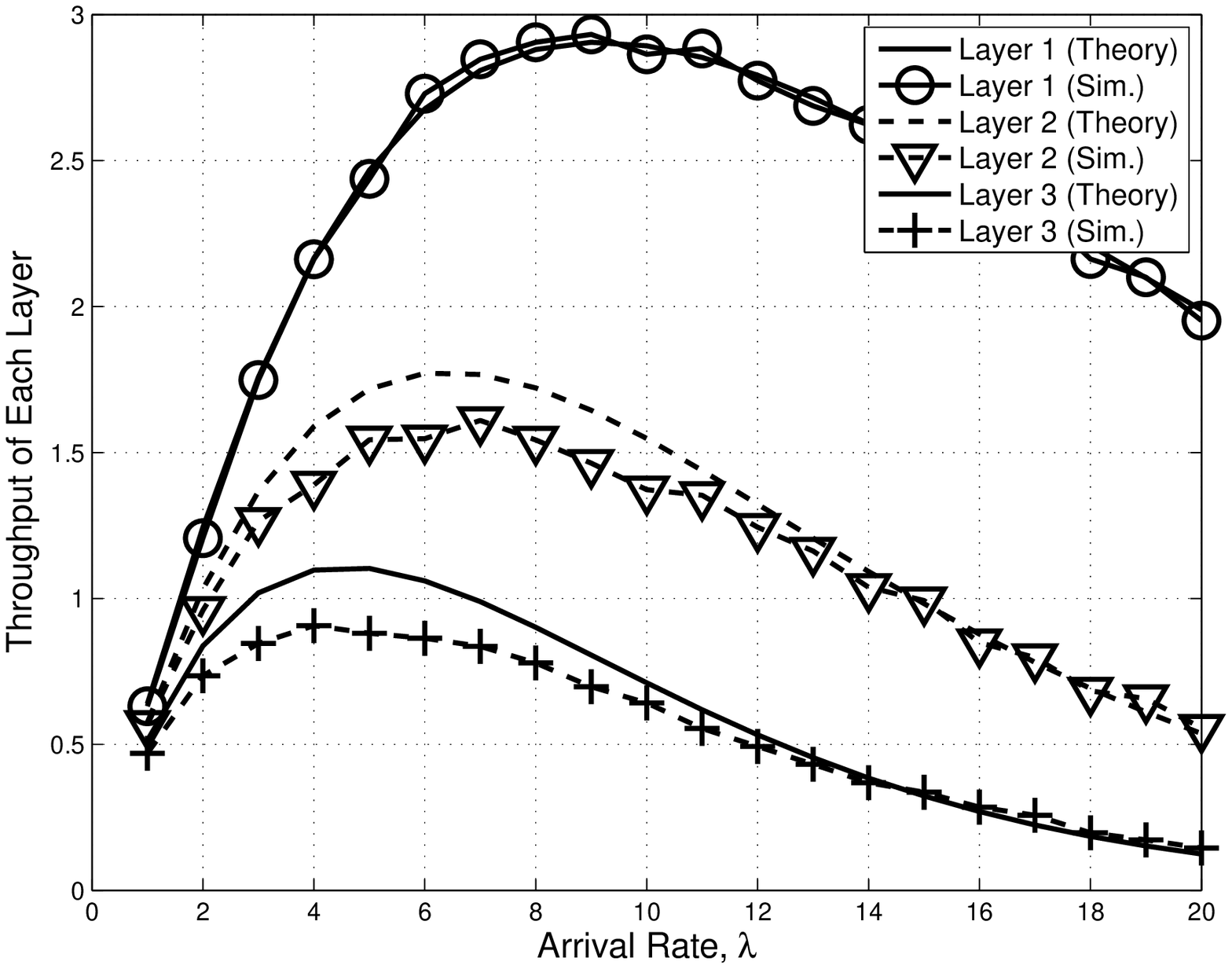} \\
(a) \\
\includegraphics[width=\figwidth]{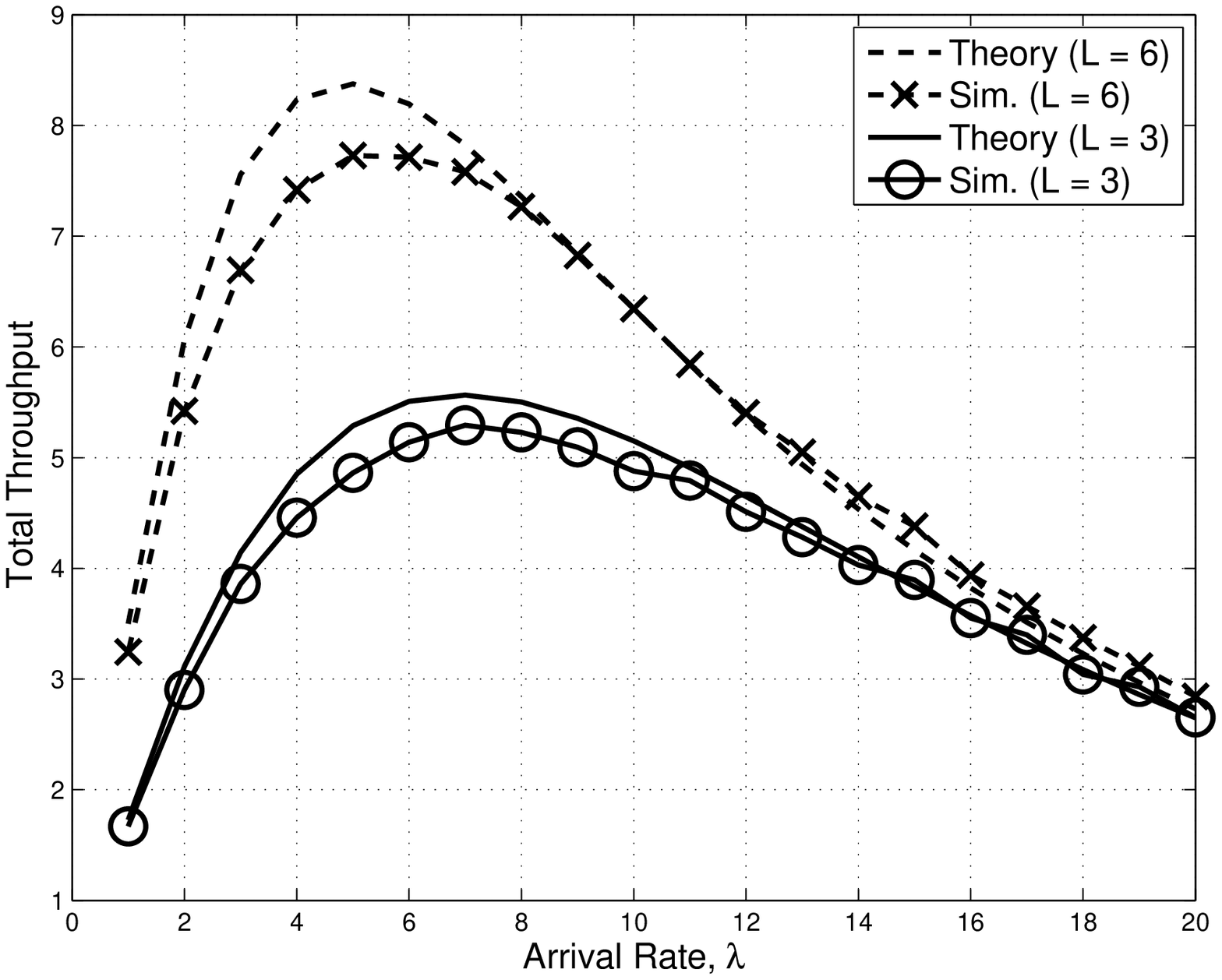} \\
(b) \\
\end{center}
\caption{Throughput versus arrival rate $\lambda$
with $\gamma = 3$ dB and $N = 10$: (a) throughput of each layer;
(b) total throughput with $L = 3$ and $L = 6$.}
        \label{Fig:plt1}
\end{figure}

Fig.~\ref{Fig:plt1}
presents the throughput of the layered random access
scheme for different arrival rate, $\lambda$,
when $N = 10$ and $\gamma = 3$ dB.
In 
Fig.~\ref{Fig:plt1} (a), with $L = 3$, the 
throughput of each layer is shown with the 
optimal rates that are found from \eqref{EQ:Rest}
for each value of $\lambda$.
We can see that the theoretical result agrees
with the simulation result for layer $1$. 
However, we find that
the theoretical result becomes an approximation
for the bottom layers, i.e., layers 2 and 3
due to the correlation of the events of successful
decoding in different layers as explained earlier.
From this, the theoretical total throughput becomes
an approximation as in Fig.~\ref{Fig:plt1} (b),
where we can also observe that the total throughput is improved
with more layers. 
In Fig.~\ref{Fig:plt1} (b),
we can also observe that the arrival rate per layer that maximizes
the throughput is less than the number of channels, $N$,
and decreases with the number of layers.
In order to avoid frequent collisions, an arrival rate lower than
$N$ might be desirable (which could result in a higher throughput).
In addition,
since the interference power increases with the number of layers,
the arrival rate can decrease 
with the number of layers for a reasonable interference power and 
to achieve a higher throughput.

\begin{figure}[thb]
\begin{center}
\includegraphics[width=\figwidth]{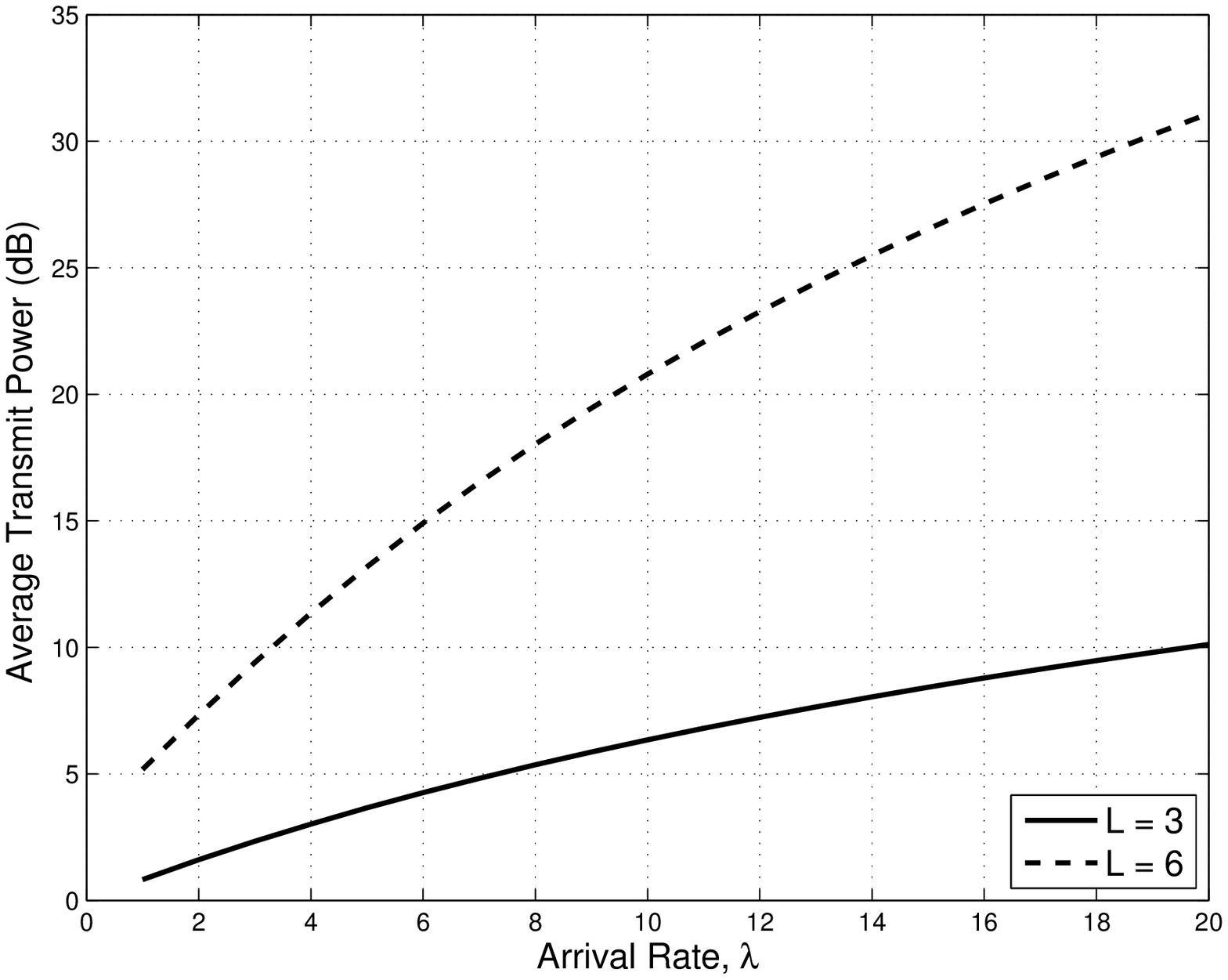} 
\end{center}
\caption{Average transmit power for different arrival rate,
$\lambda$ with $\gamma = 3$ dB,
$L \in \{3,6\}$, and $N = 10$.}
        \label{Fig:plt1c}
\end{figure}

Fig.~\ref{Fig:plt1c} shows that the transmit power increases
with the number of layers. From this,
we can see that the total throughput is improved 
at the cost of higher transmit power.
In addition, the transmit power increases with
$\lambda$ to meet the nominal
SINR because the interference power increases with $\lambda$
(which is shown in \eqref{EQ:bsl}).

\begin{figure}[thb]
\begin{center}
\includegraphics[width=\figwidth]{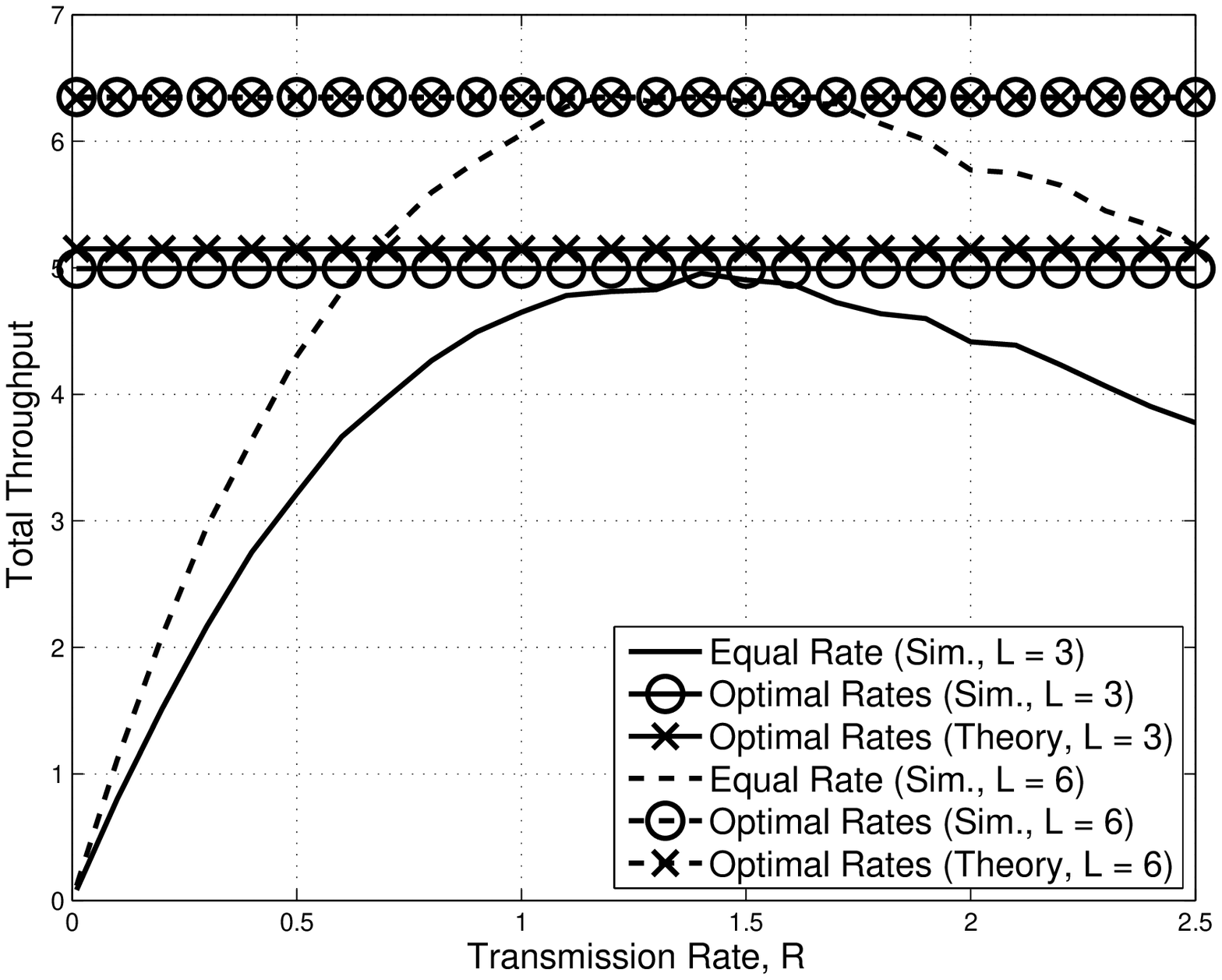}
\end{center}
\caption{Total throughput for different transmission rate, $R = R_l$,
$l = 1,\ldots, L$, when
$N = 10$, $\lambda = N$, $L \in \{3, 6\}$, and $\gamma = 3$~dB.}
        \label{Fig:plt4}
\end{figure}

In Fig.~\ref{Fig:plt4},
we assume an equal transmission rate for all the layers,
i.e., $R = R_l$, $l = 1, \ldots, L$, and show the total throughput
for different values of $R$ when
$N = 10$, $\lambda = N$, $L \in \{3, 6\}$, and $\gamma = 3$~dB.
We can observe that the optimal rates
that are obtained from \eqref{EQ:Rest} can provide the best performance
in terms of the total throughput.

\begin{figure}[thb]
\begin{center}
\includegraphics[width=\figwidth]{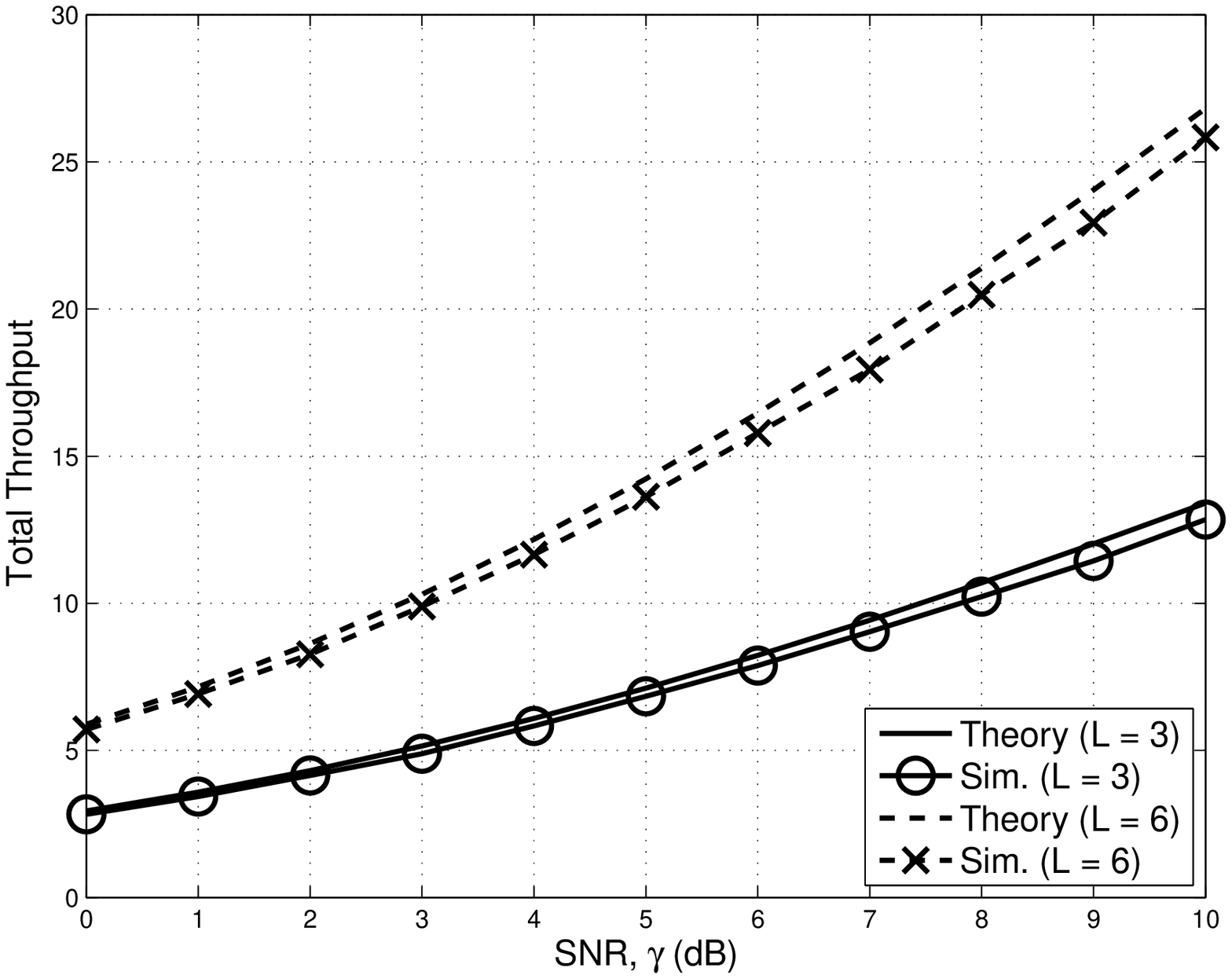}
\end{center}
\caption{Total throughput for different target SNR, $\gamma$,
when $N = 10$, $\lambda = N$, and $L \in \{3, 6\}$.}
        \label{Fig:plt2}
\end{figure}

In Fig.~\ref{Fig:plt2},
we show the total throughput 
for different target SNR, $\gamma$,
when $N = 10$, $\lambda = N$, and $L \in \{3, 6\}$.
Since $\varphi_l$ increases with $\gamma$,
the total throughput increases with $\gamma$ as shown in 
Fig.~\ref{Fig:plt2}.

\begin{figure}[thb]
\begin{center}
\includegraphics[width=\figwidth]{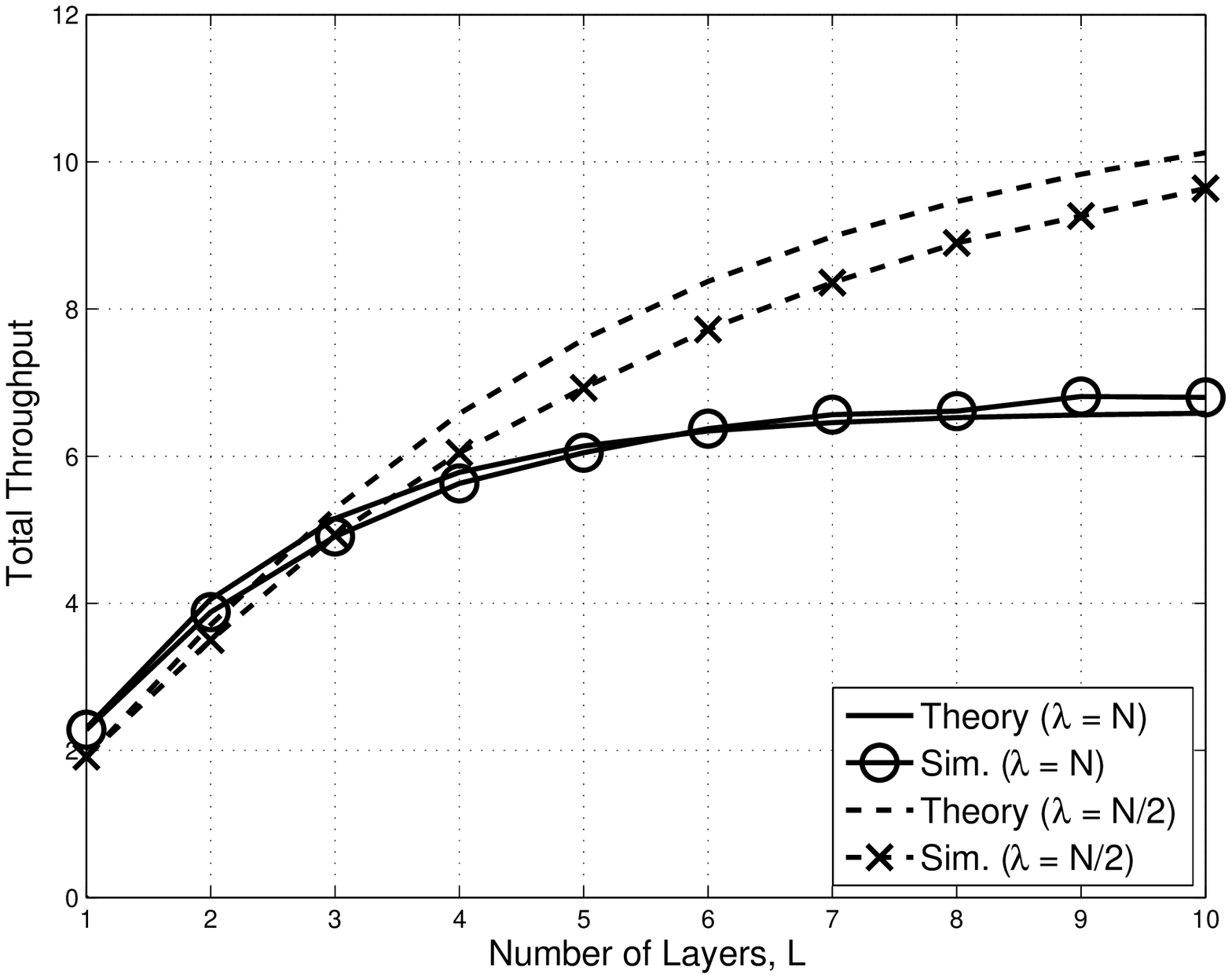}
\end{center}
\caption{Total throughput for different numbers of
layers, $L$,
when $N = 10$, $\lambda \in \{N/2, N\}$, and $\gamma = 3$~dB.}
        \label{Fig:plt3}
\end{figure}

Fig.~\ref{Fig:plt3} shows the total throughput for
different numbers of 
layers, $L$,
when $N = 10$, $\lambda \in \{N/2, N\}$, and $\gamma = 3$ dB.
The total throughput increases with $L$, while
it becomes saturated for a large $L$.
In particular, when $\lambda = N$, the 
total throughput slowly increases with $L$ when $L \ge 4$.
Interestingly, for a large
$L$ (e.g., 8), we can observe that the total throughput 
can be higher with a lower arrival rate $\lambda$.
This shows that a lower arrival rate is desirable for a larger
number of layers to keep the interference low,
which may result in a higher total throughput.

\subsection{Outage Probability}

In this subsection, we present simulation results when 
each user transmits $B$ copies of a packet through different 
channels for reliable transmissions (i.e., with CRRD). 
We show the outage probabilities
that are obtained by the theoretical approximations
(i.e., \eqref{EQ:psi_L4} and \eqref{EQ:Pel})
and simulations under the assumptions of {\bf A1},
{\bf A3}, {\bf A4}, and {\bf A5} with $\sigma_h^2 = N_0 = 1$.
Furthermore, throughout this subsection, we assume that
$R = R_l$ and $\lambda = \lambda_l$, $l = 1,\ldots,L$,
while the powers are decided as in \eqref{EQ:Pl}. 

\begin{figure}[thb]
\begin{center}
\includegraphics[width=\figwidth]{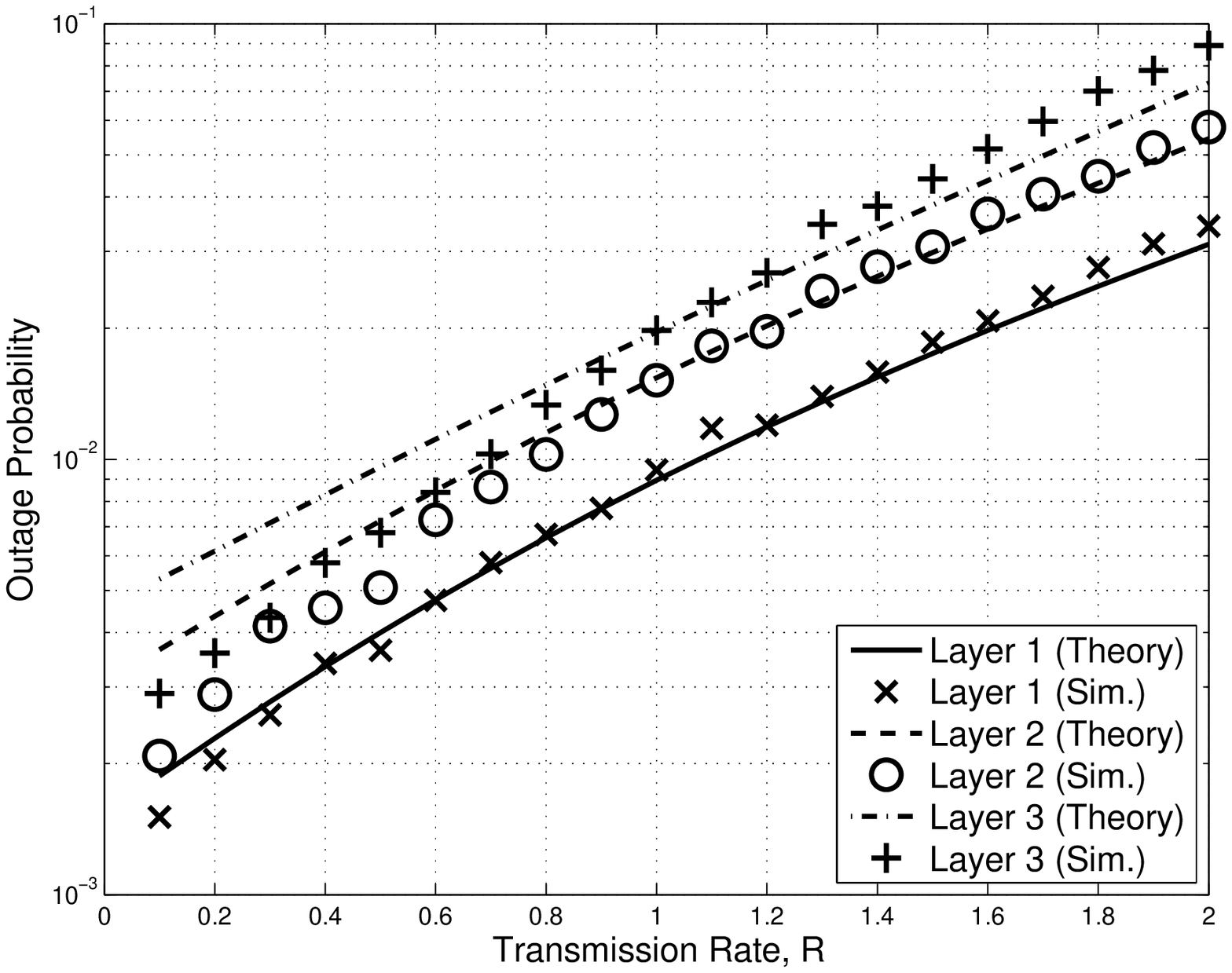}
\end{center}
\caption{Outage probabilities for different values of
transmission rate, $R$,
when $B = 4$, $N = 60$, $L = 3$, $\lambda = 3$, and $\gamma = 10$~dB.}
        \label{Fig:Bplt1}
\end{figure}

Fig.~\ref{Fig:Bplt1} shows the outage probability
for different values of transmission rate, $R$,
when $B = 4$, $N = 60$, $L = 3$, $\lambda = 3$, and $\gamma = 10$~dB.
Clearly, we can see a trade-off relationship
between the transmission rate
and reliability with a delay constraint. 
That is, we can achieve reliable transmissions
with a delay constraint at the cost of transmission rates.
We can also see that although 
\eqref{EQ:Pel} is an approximation,
it can provide reasonably good approximations of the outage
probabilities at around $R = 1$.

\begin{figure}[thb]
\begin{center}
\includegraphics[width=\figwidth]{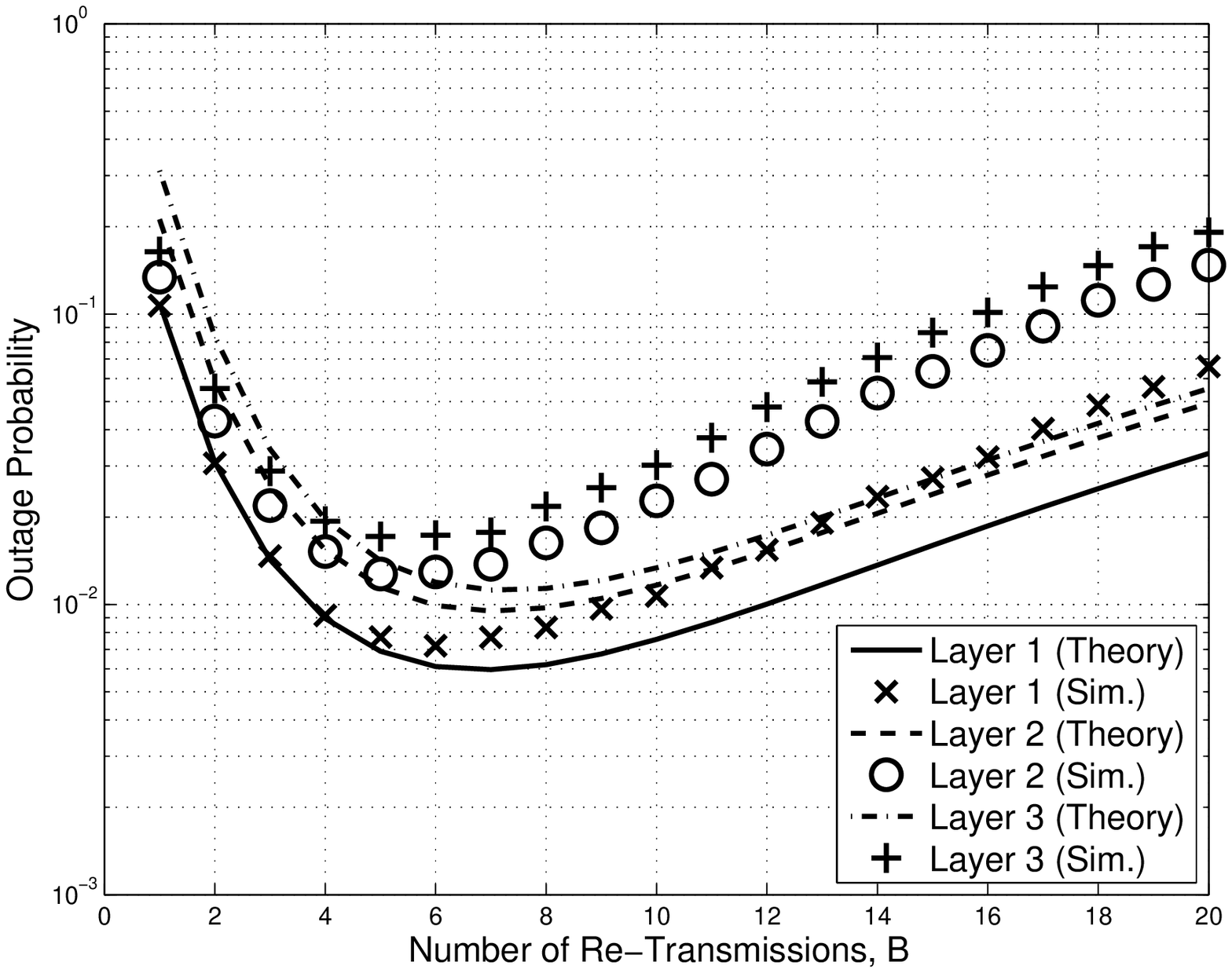}
\end{center}
\caption{Outage probabilities 
for different values of repetition gain, $B$,
when $R = 1$, $N = 60$, $L = 3$, $\lambda = 3$, and $\gamma = 10$~dB.}
        \label{Fig:Bplt2}
\end{figure}

In Fig.~\ref{Fig:Bplt2},  we present the outage probability
for different values of repetition gain, $B$,
when $R = 1$, $N = 60$, $L = 3$, $\lambda = 3$, and $\gamma = 10$~dB.
It is interesting to see that there might be an optimal $B$,
which is around 6. If $B$ is too large, there might be more collisions.
On the other hand, if $B$ is too small, the multiple transmit diversity
gain is small. Note that 
there is a noticeable gap between
the theoretical approximations and simulation results for a large
$B$, because
\eqref{EQ:psi_L4} is obtained without taking
into account the correlation of the $\sigma_{l,q}^2$'s
that increases with $B$.

\begin{figure}[thb]
\begin{center}
\includegraphics[width=\figwidth]{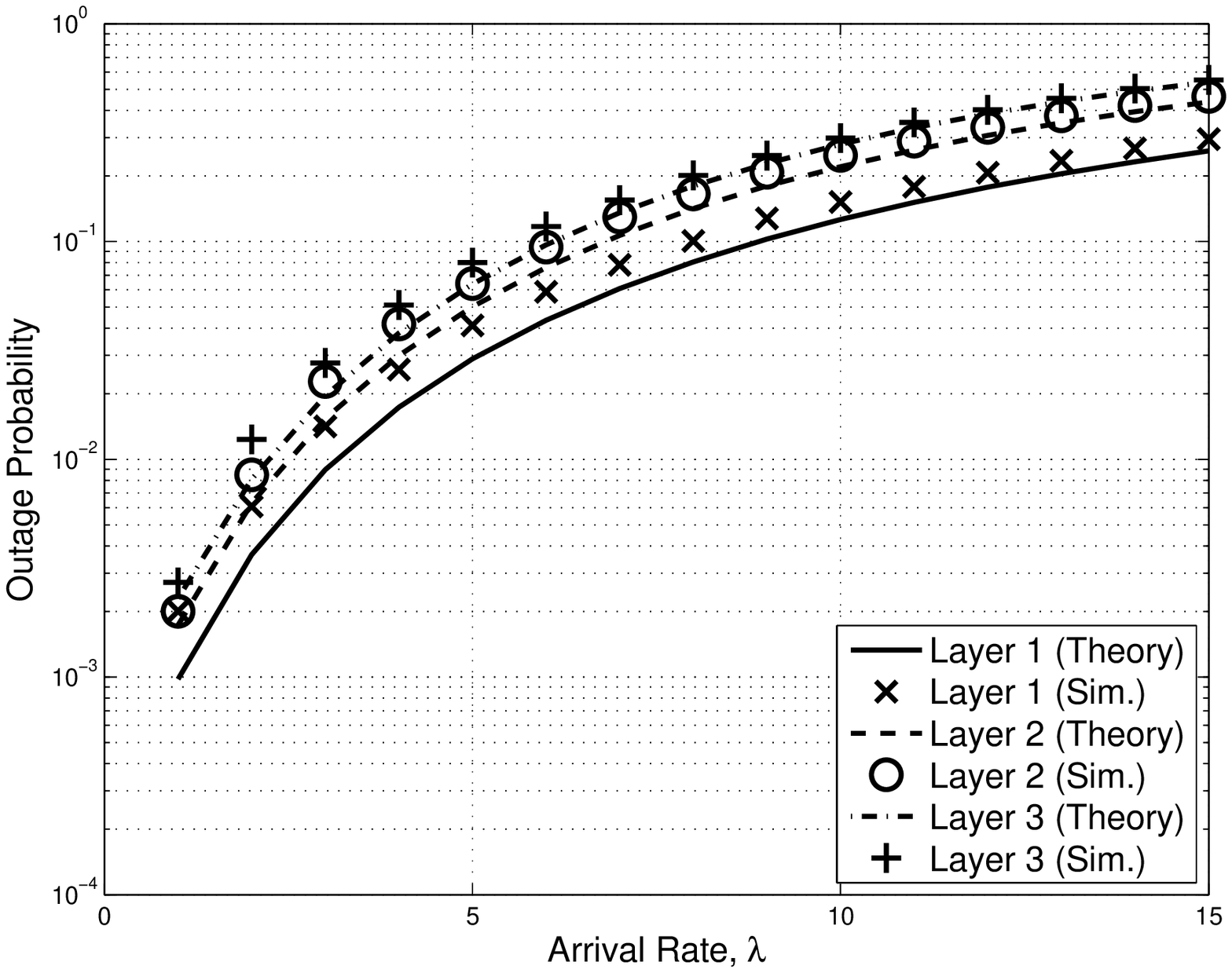}
\end{center}
\caption{Outage probabilities for different values of
arrival rate, $\lambda$,
when $B = 4$, $N = 60$, $L = 3$, $R = 1$, and $\gamma = 10$~dB.}
        \label{Fig:Bplt3}
\end{figure}

In Fig.~\ref{Fig:Bplt3},  the outage probabilities
are shown for different values of arrival rate, $\lambda$,
when $B = 4$, $N = 60$, $L = 3$, $R = 1$, and $\gamma = 10$~dB.
For a low outage probability, 
it is desirable to have a low arrival rate, $\lambda$.
This might be seen as a trade-off relationship between the throughput
and reliability with a delay constraint. 
Together with the results in Fig.~\ref{Fig:Bplt2},
we can conclude that reliable transmissions
with a delay constraint
can be achieved with CRRD at the cost of
transmission rates as well as arrival rates or throughput.

From Figs.~\ref{Fig:Bplt1} -- \ref{Fig:Bplt3},
we can see that the outage probabilities of layers are not
significantly different, while the outage probability increases with $l$
as expected.
With a large $L$, we can have more transmissions. However, the users
choosing upper layers, i.e., a large $l$, should be more tolerable for
transmission reliability.

\section{Concluding Remarks}	\label{S:Con}

We considered a layered random access scheme that can support
more users by exploiting the notion of NOMA in this paper.
To find the throughput, we derived a closed-form expression
for the probability of successful decoding by taking 
into account packet collision as well as 
decoding errors due to low instantaneous SINR. From this, 
a closed-form expression for the total throughput
was derived. Although
it is an approximation as the correlation
of the events of successful decoding
in different layers has been ignored,
it allowed us to find optimal rates that maximize the total
throughput. From simulation results, we confirmed
that the resulting optimal rates can provide the highest throughput
(as shown in Fig.~\ref{Fig:plt4}).

From brief comparisons between 
the proposed layered random access scheme and IRSA,
we found that the proposed 
layered random access scheme can provide a higher throughput using
multiple layers than IRSA. 
A generalization of 
the proposed scheme with the notion of IRSA might be an interesting topic
where a receiver can employ
not only inter-layer, but also intra-layer SIC.
This generalization might be a further research topic to be studied
in the future.

We also modified the proposed layered random 
access scheme with CRRD so that a receiver can decode
the signals from users within a slot (or MAC frame)
with a high probability.
A closed-form expression for the outage probability
is derived, which is an approximation and reasonably good
when the repetition gain is not too large.
From simulation results and analysis, we observed that
reliable transmissions can be accomplished with a high probability
at the cost of 
transmission rates as well as arrival rates or throughput.

\appendices

\section{Proof of Lemma~\ref{L:1}}	\label{A:1}

From \eqref{EQ:al}, 
we have
\begin{align}
\eta_l 
& = \sum_{m=0}^\infty m (1 - \alpha_1 (m)) P_l (m) \cr
& = (1- \beta_l) \sum_{m=0}^\infty m (1 - p_{\rm c} (m)) P_l (m),
\end{align}
where $P_l (m)$ is the probability 
that there are $m$ 
active users at layer $l$.
Under the assumption of {\bf A1}, since
$P_l (m) = \frac{\lambda_l^m e^{-\lambda_l} }{m!}$,
it follows
\begin{align}
\eta_l 
& = (1 - \beta_l) \lambda_l 
\sum_{m=1}^\infty \left(1 - \frac{1}{N} \right)^{m-1} 
 \frac{\lambda_l^{m-1} e^{-\lambda_l} }{(m-1)!} \cr
& = \varphi_l \lambda_l e^{\lambda_l \left(1- \frac{1}{N}\right)}
e^{-\lambda_l},
\end{align}
which leads to \eqref{EQ:eta_l}.
This completes the proof.

\section{Proof of Lemma~\ref{L:2}}	\label{A:2}
Under the assumption of {\bf A4}, $|h_{k,q}|^2$ becomes
an independent chi-squared random variable with 2 degrees of freedom.
Let $D_{l,q}$ denote the number of users transmitting signals
through channel $q$ and layer $l$.
Then, $\sigma_{l,q}^2$ can be given by
\begin{align}
\sigma_{l,q}^2 
& = \uE[ |n_{l,q}|^2\,|\, \{h_{k,q}\}] \cr
& = \sum_{i = l+1}^L P_i \sum_{k \in \cI_{i,q}} |h_{k,q}|^2 + N_0 \cr
& = \sum_{i=l+1}^{L} 
\frac{\sigma_h^2 P_i}{2} \chi_{2D_{i,q}}^2 + N_0,
	\label{EQ:slq}
\end{align}
where $\chi_{2D}^2$ represents an independent chi-squared random
variable with $2D$ degrees of freedom.
Under the assumptions of {\bf A1} and {\bf A2},
we can see that $D_{l,q}$ is an independent Poisson random
variable with mean $\frac{\lambda_l}{N}$.
Thus, it can be shown that
\be
\uE \left[\frac{\sigma_h^2 P_i}{2} \chi_{2D_{i,q}}^2 \right] 
= \frac{\sigma_h^2 P_i \lambda_i}{N}.
\ee

Under the assumption of {\bf A4}, from \eqref{EQ:beta_l},
the decoding error probability becomes
\be
\beta_l = 1 -
\uE \left[
\exp \left( - \frac{\sigma_l^2 \nu( R_l)  }{P_l \sigma_h^2}
\right) \right].
	\label{EQ:l11}
\ee
Then, using Jensen's inequality, we have
\begin{align}
\varphi_l
\ge
\exp \left( - \frac{\uE[\sigma_l^2] \nu( 2^{R_l}) }{P_l \sigma_h^2}
\right) 
=  \exp \left( - \frac{\nu(R_l)}{\gamma_l} \right),
\end{align}
which becomes the lower-bound in \eqref{EQ:1b}.

To find the exact expression, from \eqref{EQ:l11}, we have
\begin{align}
\varphi_l
& = e^{- \frac{\nu(R_l) N_0}{ P_l \sigma_h^2}}
\uE\left[ e^{
- \frac{\nu(R_l)}{P_l}
\sum_{i=l+1}^L \frac{P_i \chi_{2 D_{i,q}}^2}{2} } \right].
	\label{EQ:lo1}
\end{align}
Since $\chi_{2D}^2$ is a chi-squared random variable
(under the assumption of {\bf A4}), 
it can be shown that
\begin{align}
\uE\left[ e^{ - \frac{\nu(R_l)}{P_l}
\sum_{i=l+1}^L \frac{P_i \chi_{2 D_{i,q}}^2}{2} } \right] 
& = \prod_{i=l+1}^L 
\uE\left[ 
e^{ - \frac{\nu(R_l)}{P_l} \frac{P_i \chi_{2 D_{i,q}}^2}{2} }\right] \cr
& = \prod_{i=l+1}^L 
\uE\left[ \left(
\frac{1}{ 1+ \frac{\nu(R_l) P_i }{P_l}}\right)^{D_{i,q}} \right].
	\label{EQ:l12}
\end{align}
Now, noting that $D_{i,q}$ is
a Poisson random variable 
(under the assumptions of {\bf A1} and {\bf A2}), 
we have
\begin{align}
& \uE\left[ 
\left( \frac{1}{ 1+ \frac{\nu(R_l) P_i }{P_l}}\right)^{D_{i,q}} \right] \cr
& = \sum_{d=0}^\infty
\left( \frac{1}{ 1+ \frac{\nu(R_l) P_i }{P_l}}\right)^d
\frac{(\lambda_i/N)^d}{d!} e^{-\lambda_i/N} \cr
& = \exp
\left(
- \frac{\lambda_i}{N} \frac{\nu(R_l) P_i}{P_l + \nu(R_l) P_i}
\right).
	\label{EQ:l13}
\end{align}
Substituting 
\eqref{EQ:l13} into \eqref{EQ:l12},
we have
\be
\uE\left[ e^{ - \frac{\nu(R_l)}{P_l}
\sum_{i=l+1}^L \frac{P_i \chi_{2 D_{i,q}}^2}{2} } \right] 
= 
e^{
- \sum_{i=l+1}^L \frac{\lambda_i}{N} \frac{\nu(R_l) P_i}{P_l + \nu(R_l) P_i}
}.
	\label{EQ:l14}
\ee
Finally, 
substituting \eqref{EQ:l14} to \eqref{EQ:lo1},
we can obtain the exact expression in \eqref{EQ:1b}.

\section{Proof of Lemma~\ref{L:4}}	\label{A:4}

It can be shown that
\begin{align}
&  \sum_{m=1}^\infty \alpha_l^B(m) \bar P_l (m) \cr
& =
\sum_{m=1}^\infty 
\left( p_{\rm c} (m) +  (1-p_{\rm c} (m) ) \beta_l \right)^B
\bar P_l (m) \cr
& = \sum_{m=1}^\infty 
\sum_{b = 0}^B  \binom{B}{b}
(1 - \beta_l)^b p_{\rm c}^b (m) 
\beta_l^{B-b} \bar P_l (m) \cr
& = 
\sum_{b = 0}^B  \binom{B}{b}
(1 - \beta_l)^b \beta_l^{B-b} 
\sum_{m=1}^\infty p_{\rm c}^b (m) 
\bar P_l (m).
	\label{EQ:a_psi}
\end{align}
In order to find an expression with a sum of finite terms,
we can show that
\begin{align*}
\sum_{m=1}^\infty p_{\rm c}^b (m) \bar P_l (m)
& = \sum_{m=1}^\infty ( 1 -  \omega^{m-1} )^b P_l (m) \cr
& = \sum_{j=0}^b \binom{b}{j}
\sum_{m=1}^\infty
(-\omega^{m-1} )^{j} \bar P_l (m) \cr
& = 
\frac{e^{- \lambda_l} }{1- e^{- \lambda_l} }
\sum_{j=0}^b \binom{b}{j}
(- \omega)^{-j} (e^{\lambda_l \omega^j} - 1).
	\label{EQ:ft}
\end{align*}
Substituting \eqref{EQ:ft} into
\eqref{EQ:a_psi},
we have
\begin{align}
\Psi_l
& \approx \frac{e^{- \lambda_l} }{1- e^{- \lambda_l} }
\sum_{b = 0}^B  \binom{B}{b}
(1 - \beta_l)^b \beta_l^{B-b}  \cr
& \times
\sum_{j=0}^b \binom{b}{j}
(- \omega)^{-j} (e^{\lambda_l \omega^j} - 1),
\end{align}
which is \eqref{EQ:psi_L4}. 

Note that $\beta_l$ in \eqref{EQ:g_b}
is different from $1 - \varphi_l$ that can be obtained
from \eqref{EQ:1b} due to multiple transmissions (i.e., $B > 1$).
Under the assumptions of {\bf A1} and {\bf A5},
$D_{i,q}$ in \eqref{EQ:slq}
is a Poisson random variable with mean $\frac{\lambda_i B}{N}$
instead of $\frac{\lambda_i}{N}$, i.e.,
$D_{i,q} \sim {\rm Poiss} \left(
\frac{B\lambda_i}{N} \right)$.
Thus,
under the assumption of {\bf A4},
we have
\be
\uE\left[ e^{ - \frac{\nu(R_l)}{P_l}
\sum_{i=l+1}^L \frac{P_i \chi_{2 D_{i,q}}^2}{2} } \right] 
= 
e^{
- \sum_{i=l+1}^L \frac{\lambda_i B}{N} \frac{\nu(R_l) P_i}{P_l + \nu(R_l) P_i}
},
\ee
which is substituted into \eqref{EQ:lo1} to obtain 
$\beta_l$ in \eqref{EQ:g_b}.

\bibliographystyle{ieeetr}
\bibliography{mtc}

\end{document}